\documentstyle[12pt]{article}
 
\begin{document}

\baselineskip=14pt plus 0.2pt minus 0.2pt
\lineskip=14pt plus 0.2pt minus 0.2pt

\begin{flushright}
~~ \\ 
LA-UR-96-1596 \\
\end{flushright}

\begin{center}
\large{\bf Functional Forms for the Squeeze and the
Time-Displacement Operators} 

\vspace{0.25in}

\bigskip

 Michael Martin Nieto\footnote
{Email: mmn@pion.lanl.gov }

\vspace{0.3in}

{\it
Theoretical Division\\
Los Alamos National Laboratory\\
University of California\\
Los Alamos, New Mexico 87545, U.S.A.\\
\vspace{.25in}
and\\
\vspace{.25in}
Abteilung f\"ur Quantenphysik\\
Universit\"at Ulm\\
D-89069 Ulm, GERMANY\\}

\vspace{0.3in}

{ABSTRACT}

\end{center}

\begin{quotation}

\baselineskip=0.333in

Using Baker-Campbell-Hausdorff relations, the squeeze and 
harmonic-oscillator time-displacement operators are given in the form \\
$\exp[\delta I] \exp[\alpha (x^2)]\exp[\beta(x\partial)]
\exp[\gamma (\partial)^2]$, 
where $\alpha$, $\beta$, $\gamma$, and $\delta$ are 
explicitly determined.  Applications are discussed.

\vspace{0.25in}

\end{quotation}

\vspace{0.3in}

\newpage

\baselineskip=.33in

\section{Introduction}

In group theory and quantum optics, the use of Baker-Campbell-Hausdorff
relations \cite{bch,bch2}, to write unitary operators in more useful 
(often normal-ordered) forms, 
is a well-known technique.  For instance, the displacement operator,
\begin{equation}
D(\alpha)=\exp[\alpha a^{\dagger} - \alpha^* a]~,~~~~~~
\alpha = \alpha_1 + i \alpha_2 \equiv (x_0 + i p_0)/\sqrt{2}~,
\end{equation}
and the squeeze operator,
\begin{equation}
S(z) = \exp\left[\frac{1}{2}za^{\dagger}a^{\dagger}-\frac{1}{2}z^*aa\right]~,
 ~~~~~~  z = re^{i\phi} = z_1 + i z_2~, 
\end{equation}
can be written as
\begin{equation}
D(\alpha) 
=\exp\left[-\frac{1}{2}|\alpha|^2\right]
\exp[\alpha a^{\dagger}] \exp[-{\alpha}^* a] ~  \label{D}
\end{equation}
and 
\begin{eqnarray}
S(z) 
& = &  \exp\left[{\frac{1}{2}}e^{i\theta}(\tanh r)
a^{\dagger}a^{\dagger}\right]
\left({\frac{1}{\cosh r}}\right)^{({\small{\frac{1}{2}}+a^{\dagger}a})}
\exp\left[-{\frac{1}{2}}e^{-i\theta}(\tanh r)aa\right] \label{b} \\
& = &  \exp\left[{\frac{1}{2}}e^{i\theta}(\tanh r)
a^{\dagger}a^{\dagger}\right]
(\cosh r)^{-1/2}\sum_{n=0}^{\infty}
      \frac{({\rm sech} r -1)^n}{n!}(a^{\dagger})^n(a)^n   \nonumber \\
&~& ~~~~~~~~\times~
\exp\left[-{\frac{1}{2}}e^{-i\theta}(\tanh r)aa\right]~. \label{c}
\end{eqnarray}

However, such  transformations are not as well-known 
in $x-p$ space, in terms of the variables  
\begin{equation}
x=\frac{1}{\sqrt{2}}(a+a^{\dagger})~,~~~\partial = ip = 
\frac{1}{\sqrt{2}}(a-a^{\dagger})~,\label{adef}
\end{equation}
\begin{equation}
[a,~a^{\dagger}] = 1~, ~~~~~[x,~\partial]=-1~.
\end{equation}
Granted, the displacement 
operator is easily expressed as 
\begin{equation}
D(\alpha) = \exp[-ix_0p_0/2]\exp[ip_0x]\exp[-x_0\partial]~. \label{disp}
\end{equation}
but the squeeze operator 
\begin{equation}
S(z) = \exp[-z_1(x\partial+1/2)+iz_2(x^2+\partial^2)/2]~.  \label{sform}
\end{equation}
is not.  If it were, it
could be easily applied to any 
wave function, using the operator properties
\begin{eqnarray}
\exp[c\partial]h(x) &=& h(x+c) \\
\exp[\tau(x\partial)] h(x) &=& h(xe^{\tau}) \\
\exp[c(\partial^2)] h(x) &=& 
\frac{1}{[4\pi c]^{1/2}} \int_{-\infty}^{\infty}
\exp\left[-\frac{(y-x)^2}{4c}\right] h(y) dy ~.
\end{eqnarray}

In Section 2 we describe the BCH method to obtain such transformations.
In Section 3 we explicitly apply it to the squeeze operator, and
in Section 4
we do the same for the unitary,
harmonic-oscillator,  time-displacement operator 
\begin{equation}
T = \exp[-i(a^{\dagger}a + 1/2)] = \exp[-i(x^2 + \partial^2)/2]~.
\end{equation}
In Section 5 we give examples and indicate further work.  


\section{The Method}

There are a number of early papers that deal with this subject, 
but a 
complete description, that we follow, was given by Wei and Norman 
\cite{norman}.  It has been used widely.  (See, e.g., Refs. 
\cite{t,nkt}.)
Consider a unitary operator  
$U(t)$ in terms of the exponentiations of $I$, $x^2$,
$(x\partial)$, and $(\partial)^2$ 
times a parameter, $t$:
\begin{equation}
U(t) = \exp[t\{b_1I+b_2x^2 + b_3(x\partial)+b_4(\partial^2)\}]~,
\end{equation}
where the $b_i$ are  not functions of $x$ or $\partial$.
Set this equal to the following ordered product:
\begin{equation}
U_1(t) = \exp[\delta I]
\exp[i\alpha x^2]\exp[\beta(x\partial)]\exp[i\gamma(\partial)^2]~,
\end{equation}
where $\alpha(t)$ (not to be confused with the coherent state label 
$\alpha$), $\beta(t)$, $\gamma(t)$, and $\delta(t)$ are the functions 
to be determined. (Note that even in the case where U
contains only 
$I$, $x^2$, and $(\partial)^2$  terms, one still
needs a $(x\partial)$ term on the right since this operator is 
needed to close the algebra.)

Take the time derivative of both sides of the equality $U=U_1$, 
and then multiply on the right by $U^{\dagger}$
and $U_1^{\dagger}$, respectively. This yields 
\begin{eqnarray}
b_1I +b_2x^2 &+& b_3(x\partial)+b_4(\partial^2)  \nonumber \\
&=& [e^{2\delta}e^{-\beta}]\{i\dot\alpha x^2
+ \dot\beta\exp[i\alpha x^2](x\partial)\exp[-i\alpha x^2] \nonumber \\
&~&+\dot\gamma\exp[i\alpha x^2]\exp[\beta(x\partial)]
    (\partial^2)\exp[-\beta(x\partial)]\exp[-i\alpha x^2]
+\dot\delta\}~, \label{basic}
\end{eqnarray}
where ``dot" signifies $\frac{d}{dt}$.  
The factor $[e^{2\delta}e^{-\beta}]$ is unity by $UU^{\dagger}=1$.
Now  the main line 
operators on the right-hand side of the equation 
should be commuted to the right.  

First do this with
\begin{equation}
X = \exp[i\alpha x^2](x\partial)\exp[-i\alpha x^2]~.
\end{equation}
Using
\begin{equation}
e^ABe^{-A} = B + [A,B] + [A,[A,B]]/2 + \dots~,
\end{equation}
and 
\begin{equation}
[x^2,x\partial]=-2x^2~, ~~~~[x^2,[x^2,x\partial]]=0~, 
\end{equation}
one has 
\begin{equation}
X = (x\partial) -i2\alpha x^2~. \label{X}
\end{equation}
Similarly one has 
\begin{eqnarray}
Y &=& \exp[\beta(x\partial)]
    (\partial^2)\exp[-\beta(x\partial)] \nonumber \\
  &=&\partial^2\sum_{n+o}^{\infty}\frac{(-2\beta)^{2n}}{n!}
   = \partial^2e^{-2\beta}~. \label{Y}
\end{eqnarray}
This leaves  to calculate
\begin{eqnarray}
Z &=& \exp[i\alpha x^2]\partial^2\exp[-i\alpha x^2] \nonumber \\
  &=& -i2\alpha -i4\alpha(x\partial) -4\alpha^2(x^2) + \partial^2~,
    \label{Z}
\end{eqnarray}
where the last line comes from direct differentiation.  

Putting Eqs. (\ref{X}), (\ref{Y}), and (\ref{Z}) into Eq. (\ref{basic}), 
one has an equation with coefficients multiplying $I$, $x^2$, $(x\partial)$,
and $\partial^2$.  Because these are independent variables, that means 
we can write the coefficients multiplying each of these variables as 
a separate equation: 
\begin{eqnarray}
b_1 &=& \dot\gamma e^{-2\beta}(2\alpha) + \dot\delta~, \label{I} \\
b_2 &=& i\dot{\alpha} -i2\alpha\dot{\beta} 
    +i\dot{\gamma}e^{-2\beta}(-4\alpha^2)~,  \label{II} \\
b_3 &=& \dot{\beta} +\dot{\gamma}e^{-2\beta}(4\alpha)~, \label{III} \\
b_4 &=& i\dot{\gamma}e^{-2\beta} ~.  \label{IIII}
\end{eqnarray}

These are four first-order differential equations in four unknowns.  
They should be solved subject to the boundary conditions that 
the $b_i(0) = 0$.  Then set $t=1$ and one has the unitary operator
in product form.  (For the time-displacement case, one lets $t$ 
simply remain $t$, the time.)
 

\section{The Squeeze Operator}

For the squeeze operator one has, from Eq. (\ref{sform}), 
\begin{equation}
b_1=-\frac{z_1}{2}~, ~~~
b_2=i\frac{z_2}{2}~, ~~~
b_3=-{z_1}~, ~~~
b_4=i\frac{z_2}{2}~.
\end{equation}
Put this into Eqs. (\ref{I})-(\ref{IIII}).  With linear combinations of 
Eqs. (\ref{II}) to (\ref{IIII}) one finds 
\begin{equation}
\dot\alpha= -2z_2\alpha^2 -2z_1\alpha + z_2/2~.
\end{equation}
The solution that goes to zero at $t=0$ is
\begin{equation}
\alpha(t) = \frac{z^2}{2r}\frac{\sinh rt}{{\cal{S}}(t)}~,
\end{equation}
where
\begin{equation}
{\cal{S}}(t) = \cosh rt + \frac{z_1}{r} \sinh rt~.
\end{equation} 

Putting this result into a linear combination of 
Eqs. (\ref{III}) and (\ref{IIII}),
\begin{equation}
\dot{\beta} = -z_1 - 2z_2 \alpha~,
\end{equation}
or 
\begin{equation}
\beta = - \ln {\cal{S}}(t)~.
\end{equation}
Now Eq. (\ref{IIII}) gives us
\begin{equation}
\dot{\gamma} = \frac{z_2}{2} e^{2\beta}~,
\end{equation}
or 
\begin{equation}
\gamma =\frac{z^2}{2r}\frac{\sinh rt}{{\cal{S}}(t)}
= \alpha~.
\end{equation}
Finally, Eq. (\ref{I}) gives us 
\begin{equation}
\dot{\delta} = -\frac{z_1}{2} -2\alpha \dot{\gamma} e^{2\beta}~,
\end{equation}
or 
\begin{equation}
\delta = -\frac{1}{2} \ln{{\cal{S}}(t)} = \frac{\beta}{2}~.
\end{equation}

Therefore, setting $t=1$, we obtain the squeeze operator, 
\begin{equation}
S={\cal S}^{-1/2} \exp\left[\frac{iz_2}{2r}\frac{\sinh r}{{\cal S}}(x^2)\right]
\exp[-(\ln {\cal S})(x\partial)]
\exp\left[\frac{iz_2}{2r}\frac{\sinh r}{{\cal S}}(\partial^2)\right]~,
\end{equation}
where
\begin{equation}
{\cal{S}} = \cosh r + \frac{z_1}{r} \sinh r
= e^r \cos^2\frac{\phi}{2} +e^{-r}\sin^2\frac{\phi}{2}~. \label{squeeze}
\end{equation} 


\section{The Time-Displacement Operator}

The harmonic-oscillator time-evolution operator can now be calculated
in the same way, but with
\begin{equation}
b_1=0, ~~~
b_2=-\frac{i}{2}~, ~~~
b_3=0~, ~~~
b_4=+\frac{i}{2}~.
\end{equation}
Then the solutions are found in the same way, this yielding, respectively,
\begin{eqnarray}
\dot\alpha=  -2\alpha^2 -1/2~,~~~~~~~
\alpha(t) &=& -\frac{\tan t}{2} ~, \\
\alpha \dot{\beta} = 1/2 + \dot{\alpha}~,~~~~~~~
\beta(t) &=& - \ln (\cos t)~, \\
\dot{\gamma} =  e^{2\beta}/2~,~~~~~~~
\gamma(t) &=& [\tan t]/2~, \\
\dot{\delta} = -\alpha~,~~~~~~~
\delta(t) &=& -[\ln(\cos t)]/2~.
\end{eqnarray}

This means that the harmonic-oscillator time-displacement operator is 
\begin{equation}
T = [\cos t]^{-1/2}\exp\left[-\frac{i}{2} \tan t(x^2)\right]
\exp[-(\ln \cos t)(x\partial)]
\exp\left[\frac{i}{2}\tan t (\partial^2)\right]~.
\end{equation}
This result can be viewed as complementary to others \cite{hillary}
in the study of time-evolution.


\section{Discussion}

It is enlightening to look at specific examples.  

Using Eqs. (\ref{disp})
and (\ref{squeeze}) on the harmonic-oscillator ground state, 
\begin{equation}
\psi_0 =\pi^{-1/4}\exp[-x^2/2]~,
\end{equation}
one finds
\begin{eqnarray}
\psi_{ss} &=& D(\alpha)S(z)\psi_0 \nonumber \\
  &=& \frac{1}{\pi^{1/4}}\frac{\exp[-ix_0p_0/2]}{[{\cal S}(1 +i2\kappa)]^{1/2}}
  \exp\left[-(x-x_0)^2 \left(\frac{1}{2{\cal S}^2(1+i2\kappa)}-i\kappa\right)
    +ip_0x\right]~,
\end{eqnarray}
where
\begin{equation}
\kappa = \frac{z_2 \sinh r}{2rs}~.
\end{equation}
This is the most general squeezed state.  Setting $z$ to be real and 
positive  
yields the most commonly  studied example:
\begin{equation}
  \psi_{ss} =
  [\pi^{1/2}s]^{-1/2}\exp\left[-\frac{(x-x_0)^2}{2s^2} - ip_0x\right]~,
  \label{cs}
\end{equation}
where
\begin{equation}
s = e^r~.
\end{equation}

As a test, the time-evolution operator can be applied to the coherent 
states, which are Eq. (\ref{cs}) with $s=1$. Then one finds
\begin{eqnarray}
T\psi_{cs} &=& \frac{e^{-it/2}}{\pi^{1/4}}
   \exp\left[-\frac{1}{2}\{x-(x_0\cos t + p_0\sin t)\}^2\right]
         \exp[ix(p_0\cos t - x_0 \sin t)] \nonumber \\
&~&\exp\left[-\frac{i}{2}(x_0 \cos t +p_0\sin t)(p_0\cos t -x_0\sin t)\right]~.
\end{eqnarray}
Of course, a simpler calculation is 
possible by replacing $\alpha$ with $\alpha e^{-it}$
in the series defining the coherent states as an infinite sum of 
number states. 

But the  time-evolution operator can be  applied to 
more complicated systems,  for instance, the even and odd 
states of the harmonic oscillator. There, starting from the 
$z$ real and $p_0 = 0$ squeezed states one can calculate  \cite{eo}
\begin{equation}
\psi_{s\pm}(x,t) = T \psi_{s\pm}(x)~.
\end{equation}
One finds
a closed form expression for $\psi_{s\pm}$:
\begin{eqnarray}
\psi_{s\pm}(x,t)
&=& \left[\frac{s}{2\pi^{1/2}(1\pm e^{-x_0^2 \cos^2 t})}
\frac{s^2 \cos t -i\sin t}{s^4 \cos^2 t + \sin^2 t}\right]^{1/2}
\nonumber \\
&~& \left\{\exp\left[-\frac{(x-x_0\cos t)^2}{2}
  \left(\frac{s^2-i\tan t}{s^4\cos^2 t + \sin^2 t}\right) 
    - \frac{i}{2}(\tan t)x^2\right] \right. \nonumber \\
&~& ~~~ \left.    
  \pm \exp\left[-\frac{(x+x_0\cos t)^2}{2}
  \left(\frac{s^2-i\tan t}{s^4\cos^2 t + \sin^2 t}\right) 
    - \frac{i}{2}(\tan t)x^2\right] \right\}~.
\end{eqnarray}
(Observe that the 
terms $\exp[-i(\tan t)x^2/2]$ are necessary to cancel 
the singularities of the terms $\exp[ix^2\tan t/(2\sin^2t)]$ when 
$t$ is an odd multiple of $\pi/2$.)
This then yields an analytic description of the probability densities
as a function of $x,t$:
\begin{eqnarray}
  \rho_{s\pm}(x,t) &=& \frac{\exp[-(x^2+x_0^2\cos^2t)/d^2]}
                {\pi^{1/2}d[1\pm d\exp[-x_0^2/s^2]]}  \nonumber \\
       &~&~~\left\{\cosh\left(\frac{2xx_0(\cos t)}{d^2}\right)
            \pm \cos\left(\frac{2xx_0 \sin t}{d^2 s^2}\right)\right\}~,
\end{eqnarray}
where
\begin{equation}
d^2 = s^2 \cos^2t + \sin^2t/s^2~.
\end{equation}

Finally, it is amusing to apply the simple time-evolution operator for 
a particle in a box, 
\begin{equation}
T_0= \exp[+i(\partial)^2/2]~,
\end{equation}
to a number eigenstate of a particle in a box.  One finds
\begin{equation}
T_0 \sin (\pi nx) = \sin (\pi n x) \exp[-i\pi^2 n^2/2]~,
\end{equation}
the correct time-evolution of a number state.  Here the operator is 
repeating the calculation for a continuous set of boxes along the 
real axis.  

These techniques can be applied elsewhere, such as in obtaining the 
time-evolution operator for a system with different potentials.
For example, 
the time-dependent system \cite{drt1,drt2}
\begin{equation}
V(x,t) = g^{(2)}(t) x^2 + g^{(1)}(t) x + g^{(0)}(t) ~
\end{equation}
can be studied.


\section*{Acknnowledgements}

I thank D. R. Truax for many useful conversations, and S. l. Braunstein
for helpful comments. 
This work was supported by the U.S. Department of 
Energy and the Alexander von Humboldt Stiftung.  


\newpage

\baselineskip=.33in

\end{document}